\documentclass[prd,12pt,letterpaper,nofootinbib]{revtex4}
\usepackage{graphicx}
\usepackage[top=1in, bottom=1in, left=1in, right=1in]{geometry}
\usepackage{natbib}
\begin{document}

\title{Digital Instrumentation for the Radio Astronomy Community}

\author{
  Aaron~Parsons$^1$\footnote{phone: 510-406-4322\\email: aparsons@astron.berkeley.edu},
  Dan~Werthimer$^1$,
  Donald~Backer$^1$,
  Tim~Bastian$^3$,
  Geoffrey~Bower$^1$,
  Walter~Brisken$^4$,
  Henry~Chen$^1$,
  Adam~Deller$^4$,
  Terry~Filiba$^1$,
  Dale~Gary$^7$,
  Lincoln~Greenhill$^5$,
  David~Hawkins$^6$,
  Glenn~Jones$^6$,
  Glen~Langston$^3$,
  Joseph~Lazio$^8$,
  Joeri~van~Leeuwen$^9$,
  Daniel~Mitchell$^5$,
  Jason~Manley$^{1,10}$,
  Andrew~Siemion$^1$,
  Hayden~Kwok-Hay So$^{11}$,
  Alan~Whitney$^{12}$,
  Dave~Woody$^6$,
  Melvyn~Wright$^1$,
  Kristian~Zarb-Adami$^2$
}

\affiliation{\vspace{12pt}
$^1$University of California, Berkeley;
$^2$Oxford University;
$^3$National Radio Astronomy Observatory, Charlottesville;
$^4$National Radio Astronomy Observatory, Socorro;
$^5$Harvard-Smithsonian Center for Astrophysics;
$^6$California Institute of Technology;
$^7$New Jersey Institute of Technology;
$^8$Naval Research Laboratory;
$^9$ASTRON, Netherlands;
$^{10}$Karoo Array Telescope, South Africa;
$^{11}$University of Hong Kong;
$^{12}$Massachusetts Institute of Technology, Haystack Observatory
}
\maketitle

\vspace{-12pt}\noindent
Submitted for consideration by the Astro2010 Decadal Survey Committee
in the area of {\it TEC: Technology Development} for the
{\it RMS: Radio, Millimeter and Submillimeter from the Ground} 
Discipline Program Panel.

\section*{Abstract}
Time-to-science is an important figure of merit for digital instrumentation
serving the astronomical community.  
A digital signal processing (DSP) community is
forming that uses shared hardware development, signal processing libraries, and
instrument architectures to reduce development time of digital instrumentation 
and to improve time-to-science for a wide variety of projects.
We suggest prioritizing technological development supporting
the needs of this nascent DSP community.
After outlining several instrument classes that
are relying on digital instrumentation development 
to achieve new science
objectives, we identify key areas where technologies pertaining to
interoperability and processing flexibility will 
reduce the time, risk, and cost of developing the digital instrumentation
for radio astronomy. 
These areas represent focus points where support of general-purpose,
open-source development for a DSP community should be prioritized in
the next decade.
Contributors to such technological development may be centers of
support for this DSP community, science groups that
contribute general-purpose DSP solutions as part of their own 
instrumentation needs, or engineering groups engaging in research that may
be applied to next-generation DSP instrumentation.

\clearpage

\begin{figure}
\includegraphics[height=3.5in]{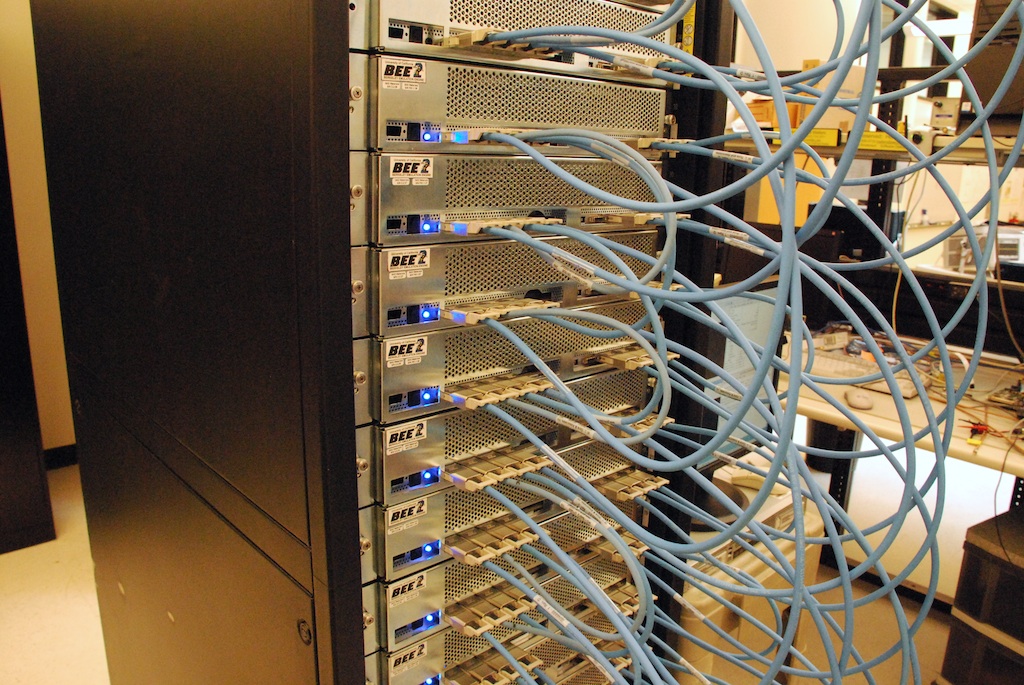}
\caption{
Cluster processing architectures using general-purpose digital hardware
and packetized communication protocols will allow a burgeoning digital
signal processing community to share development of next-generation
instruments supporting many science objectives in the coming decade.
\label{fig:bee2_rack} }
\end{figure}

\vspace{-12pt}
\section{Executive Summary}
\setcounter{page}{1}

\vspace{-12pt} 
Traditional radio astronomy signal processing instrumentation is
highly specialized; custom instruments are designed and built for individual
applications using specialized hardware, physical interconnect, communication
protocols, and control software.  In the past, custom development was required
due to project-specific constraints and the limitations of the then available
digital signal processing (DSP) technology.  Despite the explosive
growth in computational power available through DSP technology, the
complexity of development necessitates a lengthy incubation time
for individual projects, leading to a loss of timely scientific research.
To address the need for
rapid development of digital instrumentation, a ``DSP Community'' is taking
form with world-wide participantion.  This community pools the
expertise of constituent researchers and engineers around general-purpose,
open-source hardware and software resources for the timely development of new
instruments.  The evolution of this DSP Community will be critical to the
construction of new instruments and the generation of new scientific results 
over the next decade.

Pooling the resources of a diverse community of DSP 
developers requires technologies that enable hardware to be used for a variety
of applications, that enable DSP libraries to run on a variety of hardware, and
that enable instruments to be built from processors whose capabilities are
constantly growing.  We advocate for supporting the development of technologies
that commodify DSP computing, extending the concept of cluster computing
through network switches to high-performance digital processors
such as Field Programmable Gate Arrays (FPGAs, see Fig. \ref{fig:bee2_rack}), Graphics Processing Units
(GPUs), specialized DSP chips, and Application Specific Integrated Circuits
(ASICs).  The advantages of commodifying DSP computing are substantial; it
facilitates shared development of processing hardware and DSP libraries,
shortens development time for new instruments, reduces engineering costs for
maintaining and upgrading hardware, and speeds the adoption of more powerful
and energy-efficient hardware technology.  Detractors to this approach may
contend that a general-purpose system requires more resources than a system
specifically tailored to meet a particular objective; we counter that such an
approach is optimal in terms of time-to-science.

Now is an especially important time to invest in DSP technological
infrastructure.  A wide range of developing radio, millimeter, and
sub-millimeter astronomy facilities and experiments rely on
high-performance DSP computing.  These include a number of interferometric
arrays, high-bandwidth spectroscopy experiments, pulsar
de-dispersion instruments, and fast-transient searches.
Investing in technologies that catalyze
cooperative DSP instrumentation development among these projects will reduce
the total cost of achieving their many science objectives in the coming decade.
Several projects are currently demonstrating the viability of cooperative,
open-source DSP instrumentation development
\citep{parsons_et_al2006,harris_et_al2008,deller_et_al2007}, indicating that
the time is ripe for such an investment.

After describing the context of shared digital instrument development in
\S\ref{sec:motivation}, 
we outline 
in \S\ref{sec:applications} several types of
instruments that will rely upon commodification technologies to reduce
the time, risk, and cost of developing the digital instrumentation needed
to meet a
variety of science goals in the coming decade.
In \S\ref{sec:shared_dev} we discuss key areas where technological
development will be necessary to achieve many of these science goals.  These
include:
1) digitizers 2) hardware processors 3) DSP libraries 4)
flexible computing architectures 5) instrument design
6) control software.
These areas represent focus points where support of open-source 
development for a DSP community should be prioritized.

\vspace{-12pt}
\section{Motivation and Context}
\label{sec:motivation}

\vspace{-12pt}
The custom instruments that are the standard DSP solutions in
radio astronomy instrumentation usually take several years to
design, construct, and debug.  By the time they have been
deployed, their capabilities have often been surpassed by the
Moore's Law growth of computing technology.  This pattern of rapid
obsolescence is inherent to signal processing instrumentation in a
digital age and maintaining concurrency with the latest DSP
technology will be central to achieving many
science objectives in the decade ahead.  Indeed, many
projects are relying explicitly on ``just-in-time''
DSP technology development by designing instruments whose
full science objectives cannot be attained using current processors.

The capabilities of many radio astronomy applications are determined by the
availability of digital computing power and high-bandwidth interconnect.  These
applications include correlation, beam-formation, spectroscopy, 
pulsar de-dispersion, and
fast-transient searches.  The pace of DSP advances also means that radio
observatories typically need to be upgraded multiple times over their
operational lifetimes. 
The development of technologies that facilitate consistent,
scalable instrument architectures, interoperability
between families of DSP hardware, and interoperability between hardware 
generations will allow
astronomers to develop DSP libraries and instrument architectures that
easily map to new, more powerful DSP hardware as it becomes available.

We propose that priority should be given to solutions that shorten
development time for a wide range of radio astronomy DSP
applications, taking advantage of commodity hardware and interconnect
where appropriate.  For cases where new hardware is necessary, we
advocate for the development of open-source hardware
solutions that service a broad set of applications. Examples of
such hardware have already proven to be exceptionally valuable to
the radio astronomy community, and have enabled rapid progress in
a wide range of science applications \citep{parsons_et_al2006}.
Regardless of the specific DSP processors they employ, open-source
digital computing platforms enable a heterogeneous community of radio
astronomers and electrical engineers to share development costs.
The platform-independent, open-source approach reduces development
time, risk, and cost to a given project, and enhances
opportunities for innovative approaches owing to the rapid
dissemination of information and techniques.

\vspace{-12pt}
\section{Radio, MM, and Sub-mm Astronomy DSP Applications}
\label{sec:applications}

\begin{figure}\centering
\includegraphics[width=4.5in]{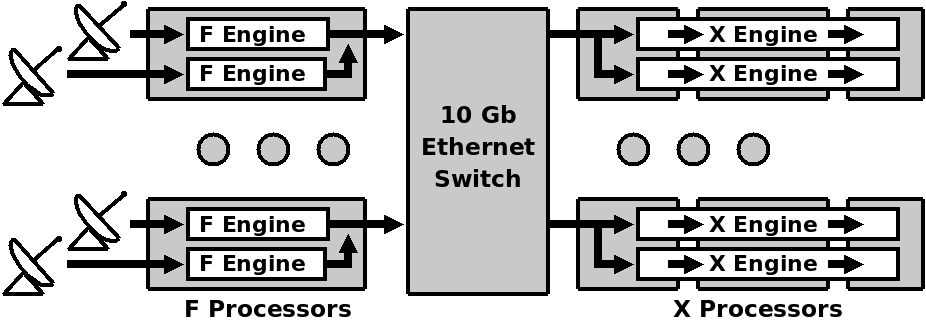}
\caption{
The CASPER packetized FX correlator architecture illustrated above
uses Field Programmable Gate Arrays
(FPGAs) connected to commercial 10-Gigabit Ethernet
switches to solve the correlator interconnect problem inherent to large
antenna arrays \citep{parsons_et_al2008}.  Using standard communication
interfaces for high-performance DSP computing is a step towards
abstracting instrument design from DSP computing hardware.
\label{fig:corr_arch}}
\end{figure}

\vspace{-12pt}
{\bf Correlators.}
Interferometric arrays use correlators to generate visibilities that
may be used for imaging.
Each cross-correlation engine in a correlator receives data from every
antenna and many engines are used to handle the aggregate
data rate; for large numbers of antennas, this easily leads to an
unmanageable number of interconnections. 
Correlator architectures that packetize antenna data can
employ commercial switches to simplify the task of routing data.
Software-based correlator architectures have often used switches
to distribute processing \citep{deller_et_al2007,west2004,takeuchi_et_al2004}.
Recently this approach has been shown to be viable for 
hardware-based correlators such as that shown in Figure \ref{fig:corr_arch},
which use 10-Gb Ethernet switches for high-bandwidth data interconnect
and fanout \citep{parsons_et_al2008}.

Two major directions of correlator development in the coming
decade will be expanding the bandwidth processed per antenna
element and expanding to large numbers of antennas and array
receivers.  Expansion
of correlator bandwidth to tens of GHz will be addressed by
developing new high-speed digitization boards (see
\S\ref{sec:shared_dev}) and parallelizing computation within
processing modules.  The overall increase in processing capability
required by higher-bandwidth correlators will most likely come in
the form of increased numbers of parallel processing modules
and should not require substantial modification of
existing correlator architectures.  However, expanding correlators
to the large numbers ($10^2$ to $10^6$) of antenna elements
required by upcoming radio astronomy facilities 
(the Allen Telescope Array (ATA), the Combined Array for Research in
Millimeter-wave Astronomy (CARMA), the Frequency Agile Solar Radiotelescope
(FASR), the Long Wavelength Array (LWA), the Murchison Widefield Array (MWA),
the Precision Array for Probing the Epoch of Reionization (PAPER), the Square
Kilometer Array (SKA))
will be a daunting task.  Developing FPGA-based
packet-switched correlators has been an important step in
demonstrating the feasibility of using high-performance switching
technology to address this problem.  However, there are still
several orders of magnitude in data routing complexity that must
be addressed before facilities of the scope of the SKA will be
feasible.

{\bf Spectroscopy and Beam-Formation.} 
Pulsar
science, solar science, and exploration of other fast-transient
sources will require versatile wideband spectrometers capable of
very rapid readout.  Flexible, high-performance
computational elements will be needed to support post-processing
tasks such as de-dispersion and RFI rejection. These applications
are well suited to hybrid DSP architectures in which digitization
and coarse channelization are performed using high-speed
ADCs mated with FPGAs, while more intricate algorithms are
implemented using CPUs and GPUs.  Other
applications such as spectral line studies and SETI will require
very high-resolution instruments. ``Zoom-in'' capability can also
be achieved with a hierarchical system where coarse channelized
data are fed to additional computational resources for
high-resolution spectroscopy.

Very Long Baseline Interferometry (VLBI) is also benefiting greatly from
advances in digital processing hardware. 
With less than two years of development, filter-banks developed on
general-purpose FPGA hardware have been combined with recently developed high
data-rate VLBI recording systems to improve the sensitivity of VLBI
observations at 1mm wavelengths (230GHz) by a factor of three, allowing
stringent new limits to be placed on the size of the presumed black hole at the
center of our galaxy \citep{doeleman_et_al2008}. Work is now proceeding to use
the same hardware platform to phase all of the mm-wavelength apertures on Mauna
Kea, enabling another large improvement in these sensitivity-starved
measurements.

{\bf Real-Time Imaging and Calibration.} For next-generation wide-field arrays,
calibrating and imaging the correlator output poses a substantial
computational burden.  At low frequencies, the need to resolve time-variable
ionospheric conditions is driving correlators to shorter integration times that
increase data rates.  For large arrays, these data rates can
reach levels where the traditional data reduction path of data storage and
off-line post-processing is no longer viable.  Heightened
time-dependent calibration requirements, wide-field
imaging with non-coplanar arrays, and real-time
RFI mitigation techniques increase the computational complexity of
post-processing.  As a result, many upcoming instruments are finding that
calibration and imaging will require digital processing comparable in
complexity to correlators \citep{wright2005}, while the algorithmic 
complexity of
real-time imaging and calibration suggests that CPU- or GPU-based
cluster-computing solutions may be appropriate \citep{wayth_et_al2007}.  

{\bf Fast-Transient Detection and Timing Systems.}
Contemporary pulsar machines take advantage of commodity CPU clusters by
channelizing the observed bandwidth into sub-bands appropriate for a single
node \citep{demorest_et_al2004}. However, such machines often rely on 
legacy high bandwidth I/O interfaces that
have quickly become obsolete. Future machines will benefit from scalable,
packetized communication between the channelizing front-end and the
computing cluster
\citep{duplain_et_al2008}. This will allow
machines to be rapidly upgraded as faster processors become available.
GPUs
have recently been shown to be very effective for coherent de-dispersion pulsar
processing, and can be rapidly added to a generic cluster to dramatically
increase performance \citep{cognard_et_al2008}.

While fast-transient radio astronomy remains a relatively unexplored field,
this is likely to change as large arrays come online. Fast-transient
observations naturally benefit from the widest bandwidth measurements possible,
because there is only one opportunity to observe any given event. This requires
a sensitive trigger to store the high-bandwidth data around an event. Flexible,
open-source DSP hardware and software will soon enable real-time searches for
dispersed fast-transient events, which can then be stored for further
processing offline.  Fast-transient processing in interferometric arrays also
benefits from a hybrid computing model involving full correlation of all
elements and processing signal from each antenna independently as they point
in different directions \citep{siemion_et_al2008}.  Fast-transient processing
can also make use of complex monitor and control systems that 
generate and receive real-time triggers for initiating follow-up
observations between observatories.

\vspace{-12pt}
\section{Shared Digital Instrumentation Development}
\label{sec:shared_dev}

\vspace{-12pt}
The variety of applications that depend on DSP
instrumentation and the unique science objectives of these
applications ensure that every DSP instrument
will be unique.  Rather than adopt a ``one-size-fits-all''
approach to shared digital instrumentation, we 
advocate developing flexible building-blocks and
architectures that allow a wide variety of
instruments with various of capabilities to be constructed.
Specialized FPGA- or ASIC-based hardware may not be
optimal for lower-bandwidth instruments or applications
that switch quickly between processing algorithms.
Similarly, high-bandwidth instruments employing relatively
simple processing algorithms may be more efficiently
implemented on such streamlined processors.

We identify six points of commonality between the various applications
discussed in \S\ref{sec:applications} where the DSP community
stands to benefit the most from shared development.  These points of
commonality include digitization hardware, digital processing hardware, DSP
libraries targeting various hardware platforms, switch-based processing
architectures, top-level instrument design, and monitor/control/interfacing
software.  Each of these points is discussed in greater detail below.

{\bf Interchangeable Digitizers.}
Designing and calibrating analog-to-digital converters (ADCs) is expensive 
and time-consuming.
Boards that employ commercially developed ADCs have typically been custom-built
for individual applications and have employed custom interfaces for passing
data to digital processors.  Currently, digitizer
boards often go through several stages of redesign as crosstalk and reflection
artifacts
are identified and eliminated.  Furthermore, the expertise
required to design such boards is rising as the signal bandwidth to be
digitized increases.  Substantial engineering time can be saved if 
1) digitizer boards
are developed cooperatively to serve many applications and 2) a standard
interface between digitizer boards and digital instrumentation is established,
so that digitizer boards may be interchangeably attached to the same DSP
engines.

One technology that will serve a large number of
upcoming low-frequency arrays is the design of low- to moderate-bandwidth
(100MHz to 500MHz) digitizers with attention to manufacturability and 
cost such that digitizers may be produced in quantities that address the
needs of arrays with many ($10^3$ to $10^6$) receivers.
Another direction in ADC technology would address the needs of
high-bandwidth applications.  In traditional wide-band instrumentation,
broad-band signals are broken up into smaller sub-bands (typically 0.5
GHz) before digitizing. The analog mixing and filtering used to generate these
sub-bands can contribute up to one third of the total cost of a backend.
Moreover, imperfect analog filtering introduces calibration errors.  With
increasing digitizer bandwidth, such systems can be replaced by digitizers operating
at intermediate-frequency (IF) or radio-frequency (RF) bandwidths, with
sub-bands extracted digitally. The increased bandwidths of
next-generation (20 to 80 Gsamples per second) digitizers will substantially
simplify analog front-ends and will make possible
new science based around wide instantaneous bandwidths.

\begin{figure}\centering
\includegraphics[width=4in]{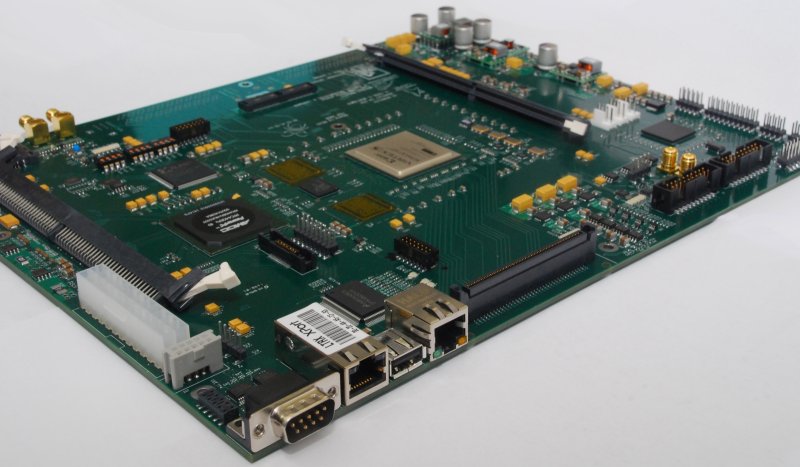}
\caption{
This example of open-source digital processing hardware was co-developed
by the Berkeley CASPER group, MeerKAT 
(the South African SKA prototype), 
and NRAO.  It features an FPGA processor that is programmable with an
open-source DSP library, and provides high-bandwidth 10-Gigabit Ethernet
interfaces for transmitting and receiving packetized data.
\label{fig:shared_hardware}}
\end{figure}

{\bf Flexible Digital Hardware Processors.}
Digital processing hardware 
runs the spectrum from lower-bandwidth,
commercially-developed CPU-based computing clusters 
to high-bandwidth, custom-developed ASICs.
Lying between these extremes are GPUs that are optimized for floating-point
operations, 
DSP-optimized microprocessors, and FPGAs that efficiently implement 
fixed-point processing.
These various processors offer a trade-off between
processing performance and programming flexibility: CPUs are programmed with
code that is fully reusable between processor generations; GPUs provide better
floating-point performance, but require code that is more customized to
architectures that change between processor generations; FPGAs have this same
trade-off, but for fixed-point operations; ASICs offer minimal programmability.
Depending on the processing needs of an application, each of these platforms
can be appropriate
\citep{deller_et_al2007,harris_et_al2008,parsons_et_al2008,ord_et_al2009}.

Currently, industry can be relied upon for developing general-purpose 
CPU and GPU
platforms that serve the needs of the DSP community. Other platforms require
custom hardware that targets the needs of radio astronomy signal processing.
General-purpose DSP hardware such as the board shown in Figure
\ref{fig:shared_hardware} demonstrate that the high development cost of these
boards can be shared between many applications \citep{parsons_et_al2006}.  Such
open-source hardware is a new direction in shared DSP instrumentation
development and represents an important step toward
commodifying high-performance DSP
processing.  A technological goal for the
next decade is designing hardware 
that employs the latest processing technology and
high-bandwidth packetized communication interfaces to function as nodes in a
cluster architecture.  As with CPU-clusters, such systems might dynamically
partition computing tasks across DSP nodes. This would allow large computing 
tasks to
be mapped into arrays of commodity processing hardware, with
heterogeneous clusters allowing DSP applications to be implemented on
platforms most suited to their performance and programmability needs.

Power consumption per operation is a processing consideration that is becoming
increasingly important for large DSP instrumentation projects.  The cost of
power and cooling for digital processing can represent a significant fraction
of the operation budget of an instrument.  The power efficiency of processors
improves with increasing density in silicon manufacturing, creating an
incentive to upgrade digital instrumentation to newer technology even when the
science goals are being met by current processors.  
One example of relatively
low-power processing acceleration uses GPUs \citep{ord_et_al2009}. While the power
consumption of a GPU may be twice that of a CPU, many applications can
achieve an order-of-magnitude increase in floating-point operations per second 
on GPUs over CPUs, and so GPUs provide more floating-point operations per watt.

{\bf Reusable DSP Libraries.}
Whether in the context of writing firmware for hardware processors or
writing software for CPU clusters, programmers have a variety of languages
and tools at their disposal.  In selecting between these 
languages/tools, a programmer may decide between low-level,
performance-oriented languages (C, VHDL, Verilog) and higher-level languages
(Python, Simulink) that trade performance for ease-of-use and flexibility.
Hybrid approaches are gaining some momentum in the software community, where
software frameworks are implemented in a high-level language and performance
bottlenecks within this framework are re-coded in a performance-oriented
language \citep{greenfield_et_al2002,smirnov_noordam2006,mcmullin_et_al2007}.  

Both performance and flexibility are priorities for libraries that are to be
shared by the DSP community.  Tools that facilitate
hybrid programming approaches 
should conceal chip-level or board-level details from programmers, establishing
a top-level interface that is abstracted from details of
hardware implementation.  As has been demonstrated by examples in the software
community \citep{haglund_et_al2003}, high-level programming interfaces increase
productivity both for casual and expert programmers.  Nonetheless, 
sometimes performance requirements can only be met using low-level
languages.  Writing cores in low-level languages requires substantial
expertise and implementations often target one generation of hardware for
a given DSP platform.  The recurring engineering cost of such cores make them
appealing targets for shared, open-source development.

The advantages of open-source software scarcely need emphasizing.
In the context of DSP instrumentation, the
advantages of shared development of DSP libraries are even more pronounced
when one
considers the necessity of porting
libraries for each hardware generation.  The quality assurance resulting from
shared development and testing for large numbers of projects far exceeds what
can be achieved for a single project.  Attention should be given in the
coming decade to implementing such DSP libraries in
open-source languages to facilitate their adoption within the community and to
ensure that all developers may easily obtain the tools the need.

{\bf Expandable and Flexible Instrument Architectures.}
The keystone technology that enables the shared development of DSP
instrumentation is the ability to combine a small set of processing
modules to create instruments that meet the needs of a wide variety of
applications.  Communication protocols and interfaces that
facilitate the interoperability of hardware modules are vital for
this goal, as are
technologies for communicating between a large number
of processing modules.  The value of these technologies have been demonstrated
for a current generation of hardware using 10-Gigabit Ethernet communication
protocols and switches.  New generations of instruments may need to make use
of other, higher-bandwidth communication solutions.  Historically,
packetized communication protocols like Ethernet tend to survive 
several hardware generations and
provide a relatively stable site for interoperability.  

The heavy reliance of internet technologies on standardized communication 
protocols ensures
the continued development of robust, high-performance switching solutions.  As
the complexity of DSP instruments increases, the advantages of employing
standardized communication become more pronounced, even when protocols incur a
modest overhead in communication bandwidth and complexity. There are also
a number of side benefits to abstracting instrument architectures from
specific hardware or generations of processing technologies.  One of these is
the ability to design an instrument using one generation of processing
technology and then to switch to latest generation hardware nearer to
the time of deployment.  By doing so, one can inexpensively expand the 
capabilities of a designed instrument or for a fixed set of capabilities,
reduce power consumption.

\begin{figure}\centering
\includegraphics[height=2in]{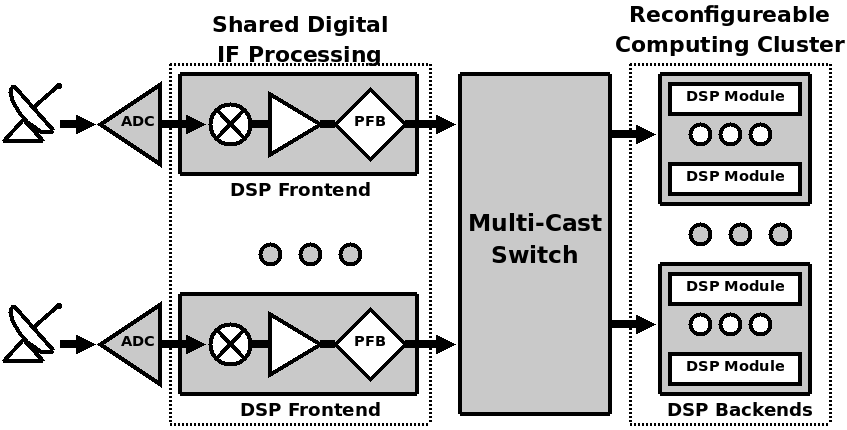}
\caption{
In a general architecture for radio astronomy DSP instrumentation, 
radio-frequency, intermediate-frequency, or
baseband signals are digitized, the relevant band is extracted, spectrally
decomposed, packetized, and transmitted in Ethernet protocol. Data are routed
through commercial multicast switches to an array of general-purpose
computing engines that can be dynamically partitioned between commensal
applications such as correlation, beam-forming, spectroscopy, pulsar
de-dispersion, and real-time imaging.  These DSP engines may employ any of
a variety of processing technologies suited to the application, including
ASICs, FPGAs, GPUs, DSP chips, and standard CPU processors.
\label{fig:general_arch}}
\end{figure}

A future direction for architecture development is to employ 
data broadcasting to promote commensal processing
by multiple backends (see Figure \ref{fig:general_arch}. 
The capability of digital processing for commensal 
observing can greatly boost the science output of radio astronomy
observatories and is a very attractive selling point of packet-switched
DSP architectures.  A related technology to be developed is the
dynamic allocation of DSP resources so that a cluster of DSP processors 
can allocate hardware computing resources much as CPU clusters do.

{\bf Shared Instrument Design.}
Although many of the design principles we have highlighted so far emphasize the
diverse nature of radio astronomy DSP applications, it is also important to
recognize the degree to which many applications overlap.  Applications
may share a fundamental architecture, even when specific design parameters
(e.g. the number of channels in spectral decomposition or the number of
antennas in an array) may differ.  Developers of independent systems can
collaborate on instrument designs that are parametrized to 
support many applications.  The modular design principles highlighted above,
ranging from interchangeable digitizers to
re-programmable processing modules to
packet-switched communication, increase the extent to which a
single parametrized instrument design can serve multiple applications.  Even
in situations where instrument designs are not explicitly 
parametrized to serve a given application, designers can
use existing, tested designs as starting points for implementing new functionality.

{\bf Shared Monitor/Control/Interface Software.}
Finally, we propose to address the boundary between
DSP instrumentation and the broader astronomcal observatory.
As signals are digitized ever closer to antenna elements and high-performance
digital processing extends deeper into backend analysis, the distinction
between DSP instrumentation and the larger observing system is becoming
vague.  Monitor, control, and interface software will be 
central to commensal digital observing, and will need to be fundamentally
integrated with the DSP systems \citep{harp_et_al2006}.

The scale and complexity of upcoming systems and their close relationship to
shared digital architectures suggest that the development of such software
might also be shared between instruments, even though monitor and control
systems must address the unique needs of each observatory.  One point where
such collaboration may be possible is the automated generation of software
drivers that interface to DSP hardware.  An example of such an interface for
FPGAs \citep{so_broderson2008} demonstrates that complex data communication
may be
abstracted using a unified file model, thereby reducing the task of remote
control and monitoring of custom DSP instruments to a familiar remote system
administration problem that is well supported by existing open-source software.

\vspace{-12pt}
\section{Summary}

\vspace{-12pt}
Time-to-science is an important figure of merit for digital instrumentation
serving the astronomical community.  The
growing capabilities of high-performance DSP computing have created the
possibility of designing hardware that serves multiple applications.  The
complexities associated with designing instrumentation have led to 
a need for pooling
expertise and development costs between multiple projects.  A DSP community is
forming that uses shared hardware development, signal processing libraries, and
instrument architectures to reduce development time of digital instrumentation 
and to improve time-to-science for a wide variety of projects.

In light of the demonstrated success of this approach and the number of
upcoming radio astronomy, millimeter, and sub-millimeter science objectives
that will rely on digital processing in the next decade, we advocate for
prioritizing the development of technologies that support
open-source, general-purpose DSP instrumentation for a broad community.
Contributors to such technological development may be centers that directly
engage in research and support for this DSP community, science groups that
offer to develop and share general-purpose DSP solutions for their own 
instrumentation needs, or engineering groups engaging in research that may
be applied to next-generation DSP instrumentation.

We have identified six areas where the DSP community is most likely to
benefit from technological development relating to processing flexibility,
standardization, and interoperability.  These are: digitization hardware,
digital hardware processors, DSP libraries, flexible
computing architectures, parametrized instrument design, and standardized
control software.  Progress in these areas will reduce the development
cost and time-to-science for DSP instrumentation at a time when a large
number of upcoming observatories and experiments will be relying on 
digital processing to
achieve their science objectives.

\clearpage
\bibliographystyle{apj}
\bibliography{biblio}

\end{document}